\documentclass[fleqn,usenatbib]{mnras}

\usepackage{newtxtext,newtxmath}

\usepackage[T1]{fontenc}

\DeclareRobustCommand{\VAN}[3]{#2}
\let\VANthebibliography\thebibliography
\def\thebibliography{\DeclareRobustCommand{\VAN}[3]{##3}\VANthebibliography}

\usepackage{graphicx}	
\usepackage{amsmath}	
\usepackage{comment}

\title[Dynamical evolution of Dimorphos boulders]{Long-term orbital evolution of Dimorphos boulders and implications on the origin of meteorites}

\author[M. Fenucci and A. Carbognani]{
M. Fenucci$^{1,2}$\thanks{E-mail: \url{marco.fenucci@ext.esa.int}  }
and A. Carbognani$^{3}$
\\
$^{1}$ESA ESRIN / PDO / NEO Coordination Centre, Largo Galileo Galilei, 1, 00044 Frascati (RM), Italy\\
$^{2}$Elecnor Deimos, Via Giuseppe Verdi, 6, 28060 San Pietro Mosezzo (NO), Italy\\
$^{3}$INAF - Osservatorio di Astrofisica e Scienza dello Spazio, Via Gobetti 93/3, 40129 Bologna, Italy\\
}

\date{Accepted XXX. Received YYY; in original form ZZZ}

\pubyear{2015}

\begin{document}
\label{firstpage}
\pagerange{\pageref{firstpage}--\pageref{lastpage}}
\maketitle

\begin{abstract}
By using recent observations of the Dydimos$-$Dimorphos system from the Hubble Space Telescope, 37 boulders with a size of 4 to 7 meters ejected from the system during the impact with the DART spacecraft were identified. In this work, we studied the orbital evolution of a swarm of boulders with a similar size to that of the detected ones. By using recent estimates for the ejection velocity of the boulders, we numerically propagated the dynamics of the swarm for 20 kyr in the future. We found that the ejection velocities and the non-gravitational effects are not strong enough to change the secular evolution significantly. The minimum orbit intersection distance (MOID) with the Earth will be reached in about 2.5 kyr, but it will not fall below 0.02 au. On the contrary, the Mars MOID will be very small in four instances, two near 6 kyr and the other two near 15 kyr. Therefore, there may be a chance for them to impact Mars in the future. Given the rarefaction of the Martian atmosphere, we expect the boulders to arrive intact on the ground and excavate a small impact crater. The results presented here provide a further indication that some meteorites found on Earth originated in collisions of $\sim$100 m near-Earth asteroids with projectiles of $\sim$1 m in size. 
\end{abstract}




\section{Introduction}
The Didymos$-$Dimorphos binary asteroid system was the subject of NASA's Double Asteroid Redirection Test \citep[DART,][]{cheng-etal_2018} mission, which aimed to test the kinetic impactor technique for asteroid deflection for the first time. (65803) Didymos is a near-Earth asteroid (NEA) of about 800 meters in size, while the moon Dimorphos has an average diameter of about 160 meters, and they form a compact binary system: the two bodies are only 1.2 km apart, and Dimorphos orbits around Didymos with a period of 11.9 hours. They are both thought to have a rubble-pile structure, which means they are not monolithic bodies but rather formed by a porous aggregation of rock boulders of different sizes \citep{walsh_2018}.    

\begin{figure}
    \centering
    \includegraphics[width=0.50\textwidth]{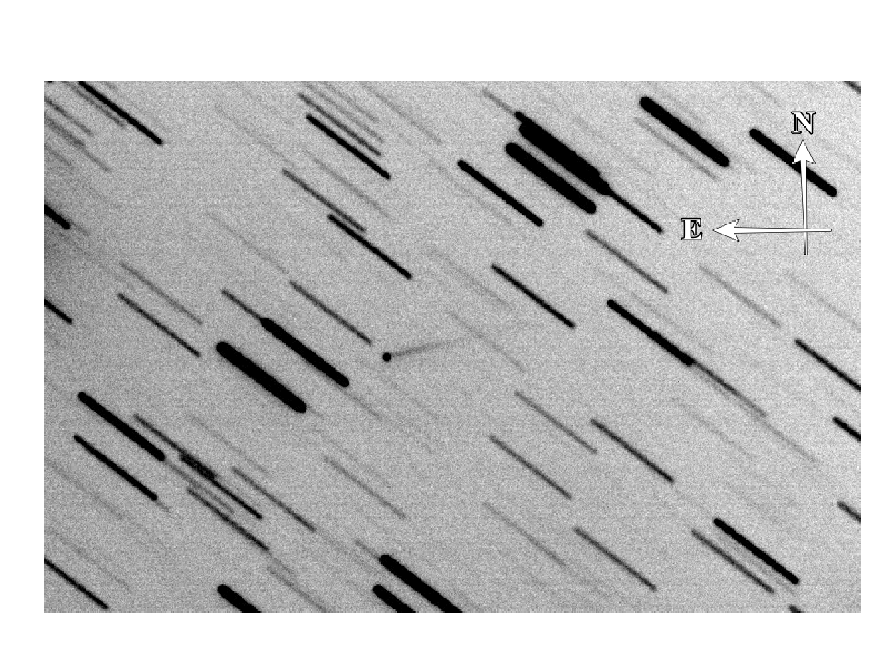}
    \caption{Image taken by AC showing the dust tail of the binary asteroid system Didymos$-$Dimorphos on 31 October 2022 with a 0.203-m telescope from the Virgil Observatory (IAU M60). In this image, the dust tail has a length of 2 arc minutes towards the position angle $275^\circ$.}
    \label{fig:Didymos_tail}
\end{figure}

The DART spacecraft successfully impacted the moon Dimorphos on 26 September 2022 \citep{daly-etal_2023}. The spacecraft also carried a small probe, the Light Italian CubeSat for Imaging of Asteroids \citep[LICIACube,][]{dotto-zinzi_2023}, that was released on 11 September 2022 and imaged the impact during its flyby with the Didymos$-$Dimorphos system. 
Ground-based observations of the event were also taken by several observatories across the world and by the Hubble Space Telescope and revealed a fast-moving plume of ejected particles produced by the impact \citep{li-etal_2023, grayakowski-etal_2023}. The dust tail remained visible for a few months after the impact, see Fig.~\ref{fig:Didymos_tail}.
Lightcurves obtained after the impact confirmed that the deflection was successful \citep{daly-etal_2023, cheng-etal_2023}, and the orbital period was reduced by about 30 minutes \citep{thomas-etal_2023}.
More details about the impact outcome and the consequences that it had on Dimorphos surface topography will be obtained with the arrival of the European Space Agency (ESA) Hera spacecraft \citep{michel-etal_2018} at the Didymos$-$Dimorphos system.

The Hubble Space Telescope was used to observe the binary system months after the impact event, specifically on 19 December 2022, 4 February 2023, and 10 April 2023. In the December image, there is a population of objects that had never been observed before: 37 boulders ejected during the collision of DART with Dimorphos are still present in the space around the binary system. These are real objects that move together with Didymos$-$Dimorphos, as demonstrated by the shooting in the following months. The apparent magnitude of the brightest objects in the December image ranges from +26.4 to +27.6 thus, assuming a geometric albedo of 0.15, the dimensions range from 4 to 7 meters, compatible with those of boulders ejected in the space due to the collision of DART \citep{jewitt-etal_2023}. The relative velocity of the 37 boulders with respect to the centre of mass of the Didymos$-$Dimorphos system, projected onto the plane of the sky, is $0.30 \pm 0.02$ m s$^{-1}$. The escape velocity of Dimorphos is only 0.09 m s$^{-1}$, while a velocity of 0.24 m s$^{-1}$ is sufficient to escape from the binary system. Therefore, all the boulders are slowly moving away from the Didymos$-$Dimorphos system, and in a certain sense, they are the slowest and easiest to observe even months after the impact.

In this work, we aim to study the long-term orbital evolution of the large boulders ejected by the DART impact. By using numerical simulations, we propagated the orbits of a swarm of boulders for 20 kyr in the future, and found that they cross the orbit of Mars. Further, we analyzed under which conditions the boulders are able to reach the surface of Mars without undergoing disruption, thus creating an impact crater.

The paper is structured as follows. In Section~\ref{s:secevol} we describe the settings of the numerical simulations of the orbital evolution and present the results. In Section~\ref{s:mars_meteorites} we present the simulations of the entry of a Dimorphos boulder in the Martian atmosphere. 
In Section~\ref{s:discussion} we discuss the results of this work and their implications, and we finally provide our conclusions in Section~\ref{s:conclusion}. 

\section{Secular evolution of the boulders}
\label{s:secevol}
Large boulders generated by the DART impact on Dimorphos were expelled with a relative velocity of $\Delta V = 0.30 \pm 0.02$ m s$^{-1}$ \citep{jewitt-etal_2023}. To generate starting conditions for the boulders, let us denote with $(\mathbf{x}, \mathbf{v})$ the initial position and velocity of an ejected particle. Assuming that Dimorphos is a sphere of 160 m in diameter, the position $\mathbf{x}$ is randomly chosen on Dimorphos surface using a uniform distribution. The initial velocity $\mathbf{v}$ is obtained by displacing Dimorphos' velocity at the impact epoch by a quantity of $\Delta V$ along the radial direction, assuming that $\Delta V$ is Gaussian distributed with an average of 0.30 m s$^{-1}$ and standard deviation of 0.02 m s$^{-1}$. The heliocentric position and velocity of Dimorphos at the time of DART's impact, needed to generate the initial conditions of the boulders, were extracted from the JPL Horizons System\footnote{\url{https://ssd.jpl.nasa.gov/horizons/}}.  To take into account the uncertainties in ejection velocity and the actual position of the impact on Dimorphos, we randomly generated a total of 3700 boulders. 

Numerical integrations are performed with the \texttt{mercury}\footnote{\url{https://github.com/Fenu24/mercury}} integrator \citep{chambers-migliorini_1997} by using the Bulirsch-Stoer scheme \citep{bulirsch-stoer_2002}. The gravitational attractions of the Sun and of all the planets from Mercury to Neptune were included in the model. In addition, we modified the code by \citet{fenucci-novakovic_2022} to take into account the effect of the solar radiation pressure (SRP). The acceleration $\mathbf{a}$ due to the SRP was modeled as
\begin{equation}
    \mathbf{a} = A_1 \frac{\mathbf{r}}{r^3}, \qquad A_1 = \frac{\phi}{c} \frac{A}{m},
    \label{eq:srp}
\end{equation}
where $\mathbf{r}$ is the heliocentric position, $r$ is the heliocentric distance, $\phi = 1361$ W m$^{-2}$ is the solar radiation energy flux at 1 au, $c$ is the speed of light, and $A/m$ is the area-to-mass ratio of the boulder \citep{montenbruck-gill_2000}. The parameter $A/m$ is computed assuming that the boulder is a sphere with a diameter of 7 m with a uniform density of $3290$ kg m$^{-3}$.
Heliocentric positions and velocities of the planets at epoch 2022-09-26 23:02:24.000 UTC, the impact time of DART on Dimorphos, were also taken from the JPL Horizons System. 
 
\begin{figure*}
    \centering
    \includegraphics[width=0.49\textwidth]{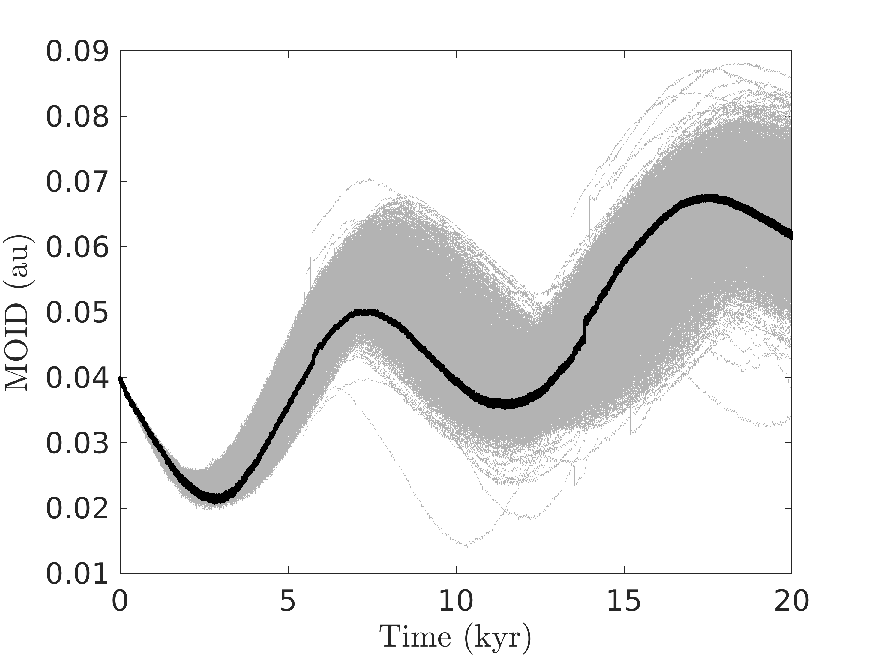}
    \includegraphics[width=0.48\textwidth]{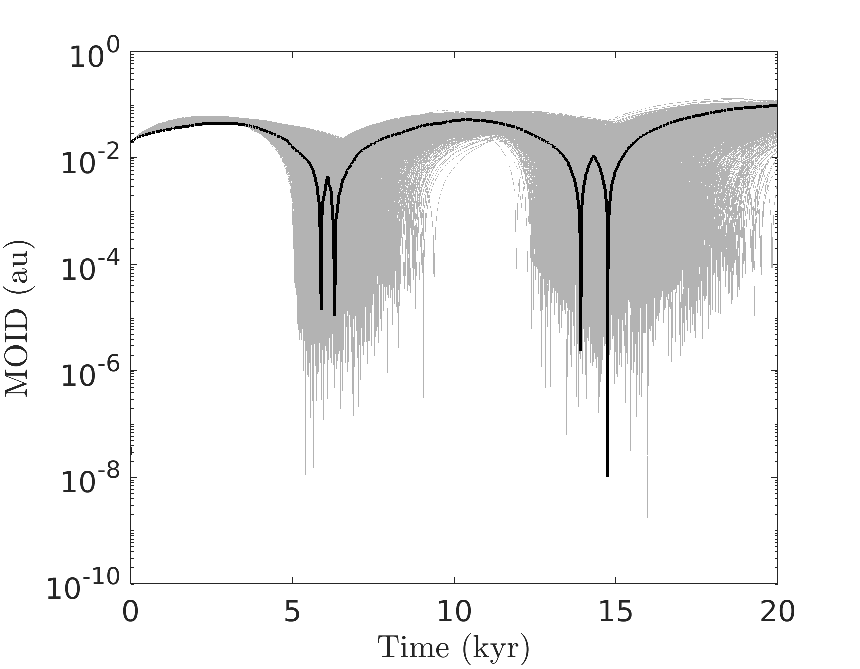}
    \caption{In grey, the time evolution of the MOID with the Earth (left panel) and with Mars (right panel), for all the 3700 generated boulders. The thick black curve is the time evolution of the MOID of the nominal orbit of (65803) Didymos.}
    \label{fig:MOID_evolution}
\end{figure*}

We numerically propagated the orbital evolution of 3700 boulders randomly generated with the model described above for 20 kyr in the future. In all the simulations, we used a maximum timestep of 2.1 h, and the Keplerian orbital elements were recorded in output every 30 days. To understand whether the boulders produced by the DART impact may produce impactors on the Earth or on Mars in the future, we computed the time evolution of the minimum orbit intersection distance (MOID) with these two planets, using the method by \citet{gronchi_2005}. The Earth and Mars MOID for all the simulated boulders are shown in Fig.~\ref{fig:MOID_evolution}. As a comparison, the evolution of the MOID of Didymos is superimposed in black. The nominal time evolution of Didymos was also simulated using the \texttt{mercury6} integrator, with initial conditions obtained from the JPL Horizons System.

The evolution of the Earth MOID of the boulders shows a minimum of $\sim$0.02 au that is reached after about 2.5 kyr. Although the MOID of the swarm of boulders shows some deviations with respect to Didymos evolution, the conditions created by the DART impact are not strong enough to significantly change the secular evolution. Although the minimum value of the MOID is somewhat smaller than the current one, it is unlikely that the boulders will fall on Earth in the next 20 kyr since 0.02 au roughly corresponds to a safe distance of about 10 lunar distances.

\begin{figure*}
    \centering
    \includegraphics[width=0.48\textwidth]{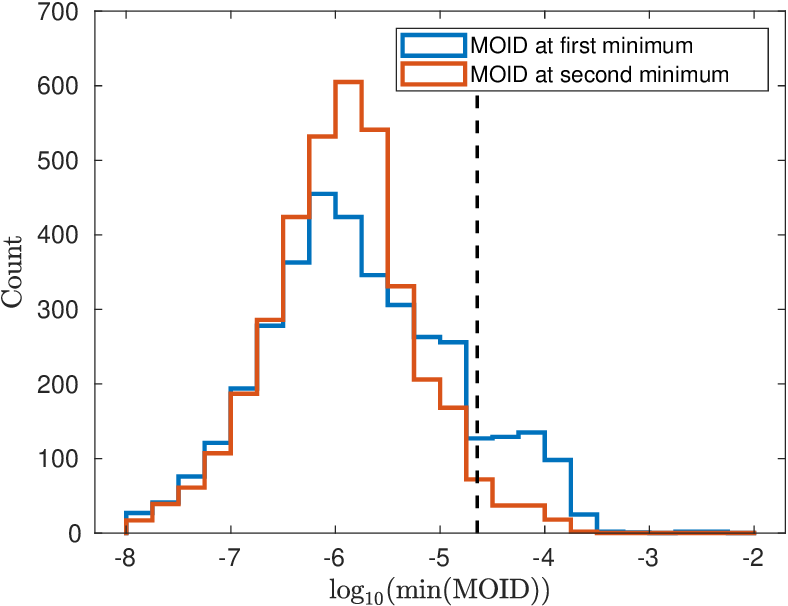}
    \includegraphics[width=0.46\textwidth]{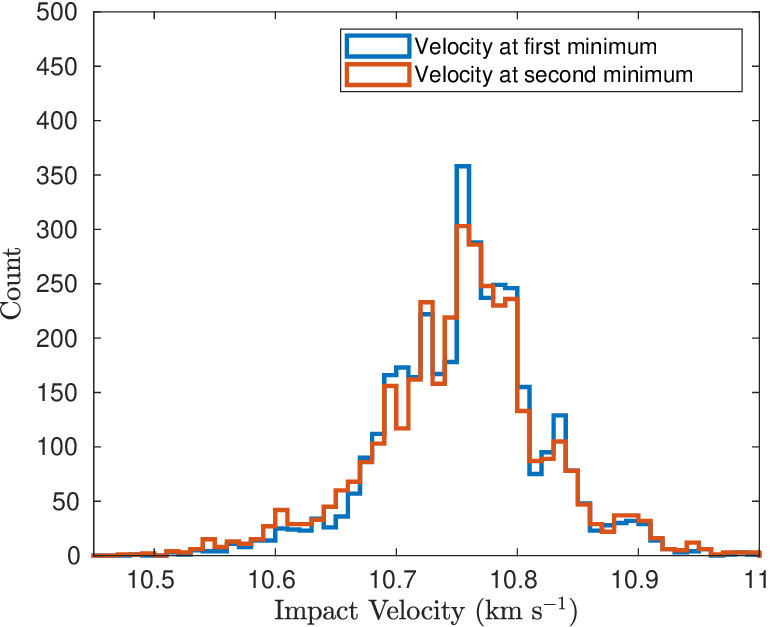}
    \caption{On the left: minimum MOID value before 8 kyr (blue histogram) and after 8 kyr (red histogram). On the right: velocity distribution at the minimum MOID at the entry in the Martian atmosphere, for the minimum before 8 kyr (blue histogram) and after 8 kyr (red histogram).}
    \label{fig:moid_speed}
\end{figure*}

On the other hand, the evolution of the Mars MOID is more noteworthy. The Mars MOID of Didymos has four minima during the 20 kyr evolution, which appear in couples at about 6 and 13 kyr. The Mars MOID of each boulder shows the same behaviour, although the times at which the minima is reached result in being spread around the epochs mentioned above. 
Figure~\ref{fig:moid_speed} shows the minimum MOID distribution reached before 8 kyr of dynamical evolution (blue histogram) and after (red histogram). The minimum MOID is smaller than the radius of Mars for 87 per cent of the boulders in the first 8 kyr and 97 per cent for the minimum MOID between 8 and 20 kyr. Overall, all the boulders have a minimum MOID smaller than the radius of Mars. Figure~\ref{fig:moid_speed} also shows the velocity distribution at the entry in the Martian atmosphere, computed at the minimum MOID. The impact velocity is always between 10.45 km s$^{-1}$ and 10.9 km s$^{-1}$. 

\begin{figure}
    \centering
    \includegraphics[width=0.48\textwidth]{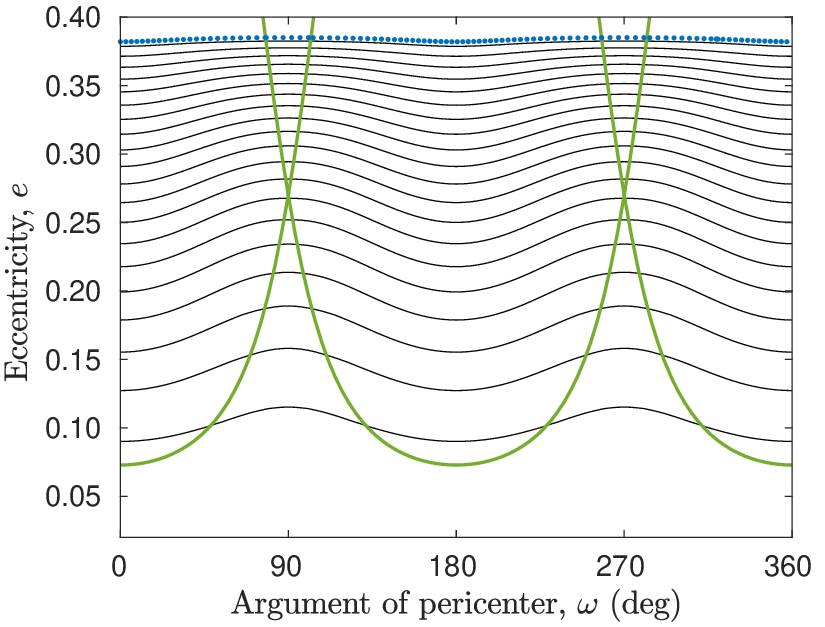}
    \caption{Secular evolution of (65803) Didymos (blue points) in the plane $(\omega, e)$. In the secular model, the orbit of an NEA follows the level curves of the secular Hamiltonian, which are shown in black. The green curves correspond to the orbit crossings with Mars.}
    \label{fig:didymos_secevol}
\end{figure}

Going more deeply into this, the fact that the orbit of Didymos crosses that of Mars on a secular timescale is known from the secular theory of the dynamics of NEAs \citep{gronchi-milani_1998, gronchi-milani_2001, fenucci-etal_2023a}. In fact, Didymos shows 4 orbit crossings with Mars\footnote{Note that this can be easily checked on the NEODyS service on the Proper Elements tab relative to Didymos: \url{https://newton.spacedys.com/neodys/index.php?pc=1.1.6&n=didymos}} in every secular cycle of $\omega$, as shown in Fig.~\ref{fig:didymos_secevol}. 
The interesting fact is that the ejection velocities of the boulders produced by the DART impact, combined with the long-term effects of the SRP, do not appear to be able to remove the crossing configurations with Mars caused by secular perturbations. Therefore, the impact of such boulders on Mars in the future can not be ruled out since they may be in the same position during the orbit crossing. However, the computation of an actual impact probability can not be performed because the perturbations in the initial orbital elements are typically accumulated in the mean anomaly in a long-term propagation; hence, predicting the exact position of the boulders along their Keplerian ellipse is practically impossible.  

\section{Possible generation of Mars meteorites}
\label{s:mars_meteorites}

In this section, we want to determine whether the boulders can reach the ground of Mars intact or if they will fragment, generating airbursts in the Martian atmosphere. As a fall model, we adopt the same one described in \cite{carbognani2024} regarding the Tunguska event based on pancake fragmentation.
For the atmosphere of Mars, we used the Viking 1 atmospheric profile obtained during entry on 20 June 1976 and provided by Planetary Data System\footnote{\url{https://atmos.nmsu.edu/data_and_services/atmospheres_data/MARS/viking/entry_profiles.html}}. The MOID atmospheric speeds of the boulders are all between 10.5 and 10.9 km s$^{-1}$ (see Fig.~\ref{fig:moid_speed}) therefore, considering that the Martian atmosphere is very thin and the atmospheric braking is reduced, we can neglect the Martian winds. Measurements of thermal inertia conducted by the OSIRIS-REx mission on the Bennu boulders have shown that there are two rock types: low and high tensile strength, with a factor of about 8 between the two categories (from 0.1 MPa to 0.8 MPa), see \cite{rozitis-etal_2020}. A strength of the same order of magnitude of the largest value, about 1 MPa, has been estimated for the Chelyabinsk fall due to an asteroid with a diameter of about 20 m and a mean density of $3290 ~\textrm{kg}~\textrm{m}^{-3}$ \citep{Borovicka2013, Popova2013}.
In our fall model for a Dimorphos boulder on Mars we assume a diameter of 7 m, an average density of $3290 ~\textrm{kg}~\textrm{m}^{-3}$, an initial atmospheric speed of 10.5 km s$^{-1}$, a starting height of 110 km, an inclination of 45° (statistically the most probable), and a drag coefficient $\Gamma \approx 0.58$. With these data, the initial kinetic energy is only 0.00778 Mt, and the mass loss due to ablation is negligible, considering that 98.6\% of the mass survived and the maximum dynamic pressure reaches a maximum of 0.96 MPa on the ground. \\
The condition of fragmentation, assuming a mean strength of about 1 MPa, is reached when the meteoroid hits the Martian soil without generating airbursts and with a speed slightly lower than the initial one because the Martian atmosphere is too thin to slow it down significantly. For this reason, in our case, a small, simple impact crater of about 200-300 m in diameter will be generated\footnote{\url{https://www.eaps.purdue.edu/impactcrater/crater_c.html}}.
However, it cannot be ruled out that Dimorphos boulders have a strength lower than about 1 MPa. In this case, the meteoroids will fragment in the Martian atmosphere without reaching the ground intact, giving rise to a classic strewn field.

\section{Discussion}
\label{s:discussion}
While the fact that the swarm of boulders crosses the orbit of Mars on a secular time-scale may not seem surprising, especially in light of the knowledge of the secular evolution of (65803) Didymos, this has, in turn, important consequences regarding the origin of meteorites. The currently accepted paradigm for the origin of meteorites is that the corresponding meteoroid originated in a collision between two main-belt asteroids, creating an asteroid family \citep[see][and references therein for an overview on asteroid families]{novakovic-etal_2022}. Later, the fragments drifted in semi-major axis because of the Yarkovsky effect \citep[see e.g.][]{bottke-etal_2006, vokrouhlicky-etal_2015}, until they reached a mean-motion resonance with Jupiter or the border of the $\nu_6$ secular resonance with Saturn. These resonances have the effect of increasing the eccentricity \citep[][]{farinella-etal_1994, bottke-etal_2002, granvik-etal_2017} moving, therefore, the fragments into the inner Solar System, where they can cross the orbit of the Earth and eventually impact the planet.  

However, some meteorites may also originate directly in the near-Earth space from NEAs since this is a dynamic population on its own. In fact, NEAs can undergo collisions, tidal disruption during flyby with a planet \citep{zhang-michel_2020}, disaggregation due to spin-up \citep{scheeres_2018}, and thermal fragmentation \citep{delbo-etal_2014, granvik-etal_2016}, thus leaving a stream of ejected material that may consequently hit the Earth and produce meteorites.
An example of this process, although the ejected material is not big enough to produce meteorite-dropping events, is represented by the Geminids meteor shower \citep{whipple_1983, williams-wu_1993}, which is caused by activity events similar to that of comets on the NEA (3200) Phaethon \citep{jewitt_2012, MacLennan2023}.

In recent years, observational data about fireballs grew dramatically thanks to the development of all-sky camera networks such as the Italian Prima Rete Italiana per la Sorveglianza sistematica di Meteore e Atmosfera\footnote{\url{http://www.prisma.inaf.it/}} \citep[PRISMA,][]{gardiol-etal_2016}, the French Fireball Recovery and InterPlanetary Observation Network\footnote{\url{https://www.fripon.org/}} \citep[FRIPON,][]{colas-etal_2020}, the European Fireball Network\footnote{\url{https://www.allsky7.net/}} \citep{oberst-etal_1998}, or the Spanish Meteor Network\footnote{\url{http://www.spmn.uji.es/}} \citep[SPMN,][]{trigo-rodriguez-etal_2001}. The large amount of data available permitted to compute the pre-atmospheric heliocentric orbit of many fireballs detected by cameras, and in recent years, several dynamical studies were performed to find associations between fireballs and meteorites with NEAs. 

In \citet{pena-asensio-etal_2022}, the authors found one fireball detected by the SPMN that could be associated with a NEA.
By further studying the fireball database of the NASA JPL Center for NEOs Studies\footnote{\url{https://cneos.jpl.nasa.gov/fireballs/}} (CNEOS), \citet{pena-asensio-etal_2023} proposed that 4\% of them may be produced by NEAs.
Regarding meteorites, \citet{delafuente-delafuente_2019} analyzed the backward orbital evolution of 2018 LA, an asteroid that impacted Earth on 2 June 2018 over Botswana and produced the Motopi Pan meteorite \citep{jenniskens-etal_2021}. The authors suggested that 2018 LA may have a common origin with the NEA  (454100) 2013 BO$_{73}$. 
By using the $D_N$ criterion \citep{valsecchi-etal_1999}, \citet{gardiol-etal_2021} proposed the association of the Cavezzo meteorite with the NEA 2013 VC$_{10}$.
More recently, \citet{carbognani-fenucci_2023} performed a larger search for parent bodies of meteorites using the $D_N$ criterion and backward numerical integrations and found 12 meteorites possibly associated with a NEA progenitor. They further suggested that about 25 per cent of meteorites may originate from collisions between $\sim$100 m sized NEAs and $\sim$1 m sized projectiles. In another study, \citet{hlobik-toth_2024} proposed 27 associations between meteorites with known orbit and NEAs, suggesting that the percentage of meteorites originating directly in the near-Earth region may be as large as 70 per cent.

The hypothesis that some meteorites may originate in collisions on small NEAs could not be tested in detail before the DART mission because the physical properties of such small asteroids were not completely known. In particular, the outcome of an impact with a $\sim$1 m sized projectile could not be fully predicted, and it was only possible to speculate on the results obtained by numerical simulations of such events \citep[see][for a review]{jutzi-etal_2015}. However, the in-situ observations of the DART impact obtained by LICIACube \citep{dotto-zinzi_2023}, together with Hubble Space Telescope observations and the ground-based taken during the event, revealed many details of the impact outcome to an outstanding level \citep{coralie_2023, cheng-etal_2023, grayakowski-etal_2023, jewitt-etal_2023, thomas-etal_2023}, which will be improved further with the arrival of the Hera spacecraft by ESA at the Didymos$-$Dimorphos system \citep{michel-etal_2018}. These results permitted us to find that boulders smaller than 10 m in diameter can be produced by such an impact and that they are ejected at a speed large enough to escape the gravitational attraction of the parent body. As we showed before, in the case of the DART impact on Dimorphos, the ejection velocity coupled with the SRP is not strong enough to significantly change the secular evolution, and the boulders still intersect the orbit of Mars as the parent body does. Therefore, there is a chance that such boulders may impact Mars in the future, and they are large enough to fly through the Martian atmosphere and reach the surface without undergoing fragmentation, thus creating a small crater. 

In the case of the Dimorphos boulders, their ejection into space is a consequence of the collision with a probe, but the same result can happen because of natural events. It is, therefore, possible that some meteorites found on Earth originated on a collision between a $\sim$100 m NEA with a projectile of $\sim$1 m in diameter as proposed by \cite{carbognani-fenucci_2023}, especially in the case the parent body was initially placed on an Earth-crossing orbit. In fact, the kinetic energy of the DART probe (mass of 570 kg, impact speed 6.6 km s$^{-1}$) is about 16 times lower than the kinetic energy of a small NEA of 1 m in diameter hitting with an average speed of about 15 km s$^{-1}$ assuming an average density of $3290 ~\textrm{kg}~\textrm{m}^{-3}$ (i.e. similar to that of the Chelyabinsk asteroid). Hence, a very small NEA of about 1 m in diameter appears capable of ejecting boulders from the surface of a 100-m diameter rubble pile asteroid.
In addition, it is important to note that, although the percentage of meteorite originating from an impact on NEAs may be high, it does not imply that we may be able to observe meteorite streams. In fact, we believe that a single-time collision would not be significant enough to create a periodic meteorite stream, as opposed to meteor showers that are created by more regular effects such as activity events.

%
 
As a final comment, all the observations taken so far prove that DART has been a successful test for asteroid deflection since it managed to change the orbital period of Dimorphos and it did not create any other boulder that could impact on Earth. On the other hand, the findings presented in this work suggest that future missions involving an interaction with the surface material of an NEA shall be carefully planned. In particular, the long-term dynamics of the ejected material shall be analyzed in detail to ease the impact monitoring on Earth and ensure the safety maintenance of operational satellites. This would be particularly important for private asteroid mining missions because NEAs orbiting closer to the Earth represent the optimal targets to maximize the cost-to-profits ratio of such missions. 

\section{Conclusions}
\label{s:conclusion}
In this work, we studied the long-term evolution of a swarm of boulders with initial conditions compatible with the ejecta produced by the DART impact on Dimorphos. Numerical simulations show that all the boulders of the swarm will cross the orbit of Mars multiple times in the future 20 kyr. 
The simulated swarm is statistically representative of the set of 37 actual boulders recently discovered by using observations from the Hubble Space Telescope that were ejected during the impact of the DART spacecraft on Dimorphos.
Therefore, due to the orbit crossings happening in the long-term evolution, it is possible that some of the boulders will impact Mars in the future. 
Simulations of the entry of a 7 m boulder from Dimorphos in the Martian atmosphere suggest that they do not undergo fragmentation, arriving intact on the ground, thus producing a small impact crater unless they have a strength lower than 1 MPa, in which case they will give rise to a strewn field.
Our results support our previous work \citep{carbognani-fenucci_2023} that some of the meteorites found on Earth may have originated collisions between $\sim$100 m NEAs with small $\sim$1 m projectiles, especially if the target was originally placed on an Earth-crossing orbit.


\section*{Data Availability}
The data underlying this article will be shared on reasonable request to the corresponding author.



\bibliographystyle{mnras}
\bibliography{holybib.bib}{}









\bsp	
\label{lastpage}
\end{document}